\documentclass{aa}
\usepackage{txfonts,graphicx,natbib}
\bibpunct{(}{)}{;}{a}{}{,}

\begin{document}

\title{Challenging GRB models through the broadband dataset of GRB\,060908}

\author{S. Covino\inst{1} \and S. Campana\inst{1}  \and M. L. Conciatore\inst{2} \and V. D'Elia\inst{3} \and E. Palazzi\inst{4} \and C. C. Th\"{o}ne\inst{1} \and S. D. Vergani\inst{5,6}  \and K. Wiersema\inst{7} \and M. Brusasca\inst{1}   \and A. Cucchiara\inst{8} \and B. E. Cobb\inst{9} \and A. Fern\'andez-Soto\inst{10} \and D. A. Kann\inst{11} \and D. Malesani\inst{12} \and N. R. Tanvir\inst{7} \and L. A. Antonelli\inst{3} \and M. Bremer\inst{13} \and A. J. Castro-Tirado\inst{14} \and A. de Ugarte Postigo\inst{1} \and E. Molinari\inst{15} \and L. Nicastro\inst{4} \and M. Stefanon\inst{16} \and V. Testa\inst{3} \and G. Tosti\inst{17} \and F. Vitali\inst{3} \and L. Amati\inst{4} \and R. Chapman\inst{18,19}  \and P. Conconi\inst{1} \and G. Cutispoto\inst{20} \and J. P. U. Fynbo\inst{12} \and P. Goldoni\inst{5,6} \and C. Henriksen\inst{21} \and K. D. Horne\inst{22} \and G. Malaspina\inst{1} \and E. J. A. Meurs\inst{23} \and E. Pian\inst{24,25} \and L. Stella\inst{3} \and G. Tagliaferri\inst{1} \and P. Ward\inst{23} \and F. M. Zerbi\inst{1}}

\institute{
INAF/Osservatorio Astronomico di Brera, via Emilio Bianchi 46, 23807, Merate (LC), Italy \and
Harvard-Smithsonian Center for Astrophysics, MA 02138, USA \and
INAF/Osservatorio Astronomico di Roma, via di Frascati 33, I-00040 Monteporzio Catone (Roma), Italy \and
INAF/Istituto di Astrofisica Spaziale e Fisica Cosmica di Bologna, via Gobetti 101, I-40129 Bologna, Italy \and
APC, Laboratoire Astroparticule et Cosmologie, UMR 7164, 11 Place Marcelin Berthelot, 75231, Paris Cedex 05, France \and
CEA Saclay, DSM/DAPNIA/Service d'Astrophysique, 91191, Gif-s\^ur-Yvette, France \and
Department of Physics and Astronomy, University of Leicester, University Road, Leicester LE1 7RH,
United Kingdom \and
Department of Astronomy \& Astrophysics, 525 Davey Lab., Pennsylvania State University, University Park, PA 16802, USA \and
Department of Astronomy, 601 Campbell Hall, University of California, Berkeley, CA 94720--3411, USA \and
Instituto de Fisica de Cantabria (CSIC-UC), Av. los Castros s/n, Santander 39005, Spain\and
Th\"uringer Landessternwarte Tautenburg, Sternwarte 5, 07778 Tautenburg, Germany \and
Dark Cosmology Centre, Niels Bohr Institute, University of Copenhagen, Juliane Maries vej 30, 2100, Copenhagen, Denmark \and
Institut de Radio Astronomie Millim\'etrique (IRAM), 300 rue de la Piscine, 38406 Saint-Martin d'H\`eres, France \and
Instituto de Astrof\'{\i}sica de Andaluc\'\i{}a, PO Box 3.004, 18080 Granada, Spain \and
INAF/TNG Fundaci\'on Galileo Galilei - Rambla Jos\'e Ana Fern\'andez P\'erez, 7, 38712, Bre\~{n}a Baja, TF - Spain\and
Observatori Astronomic de la Universitat de Val\`encia, Paterna-46980, Valencia, Spain\and
Dipartimento di Fisica e Osservatorio Astronomico, Universit\`a di Perugia, via A. Pascoli, 06123, Perugia, Italy \and
Centre for Astrophysics Research, University of Hertfordshire, College Lane, Hatfield AL10 9AB, UK \and
Centre for Astrophysics and Cosmology, Science Institute, University of Iceland, Dunhagi 5, IS-107 Reykjav'k, Iceland\and
INAF/Catania Astrophysical Observatory, via S. Sofia 78, 95123 Catania, Italy \and
Niels Bohr Institute, Blegdamsvej 17, 2100 Copenhagen, Denmark\and
SUPA Physics and Astronomy, University of St Andrews, North Haugh, St Andrews KY 9SS, Scotland, UK\and
Dunsink Observatory - DIAS, Dunsink Lane, Dublin 15, Ireland\and
INAF/Osservatorio Astronomico di Trieste, Via G. Tiepolo 11, 34143 Trieste, Italy\and
Scuola Normale Superiore, Piazza dei Cavalieri 7, 56126 Pisa, Italy
}

\date{Received <date> / Accepted <date>}

\titlerunning{The prompt and the afterglow of GRB\,060908}
\authorrunning{Covino et al.}

\abstract
{Multiwavelength observations of gamma-ray burst prompt and afterglow emission are a key tool to disentangle the various possible
emission processes and scenarios proposed to interpret the complex gamma-ray burst phenomenology.}
{We collected a large dataset on \object{GRB\,060908} in order to carry out a comprehensive analysis of the prompt emission
as well as the early and  late afterglow.}
{Data from \textit{Swift}-BAT, -XRT and -UVOT together with data from a number of different ground-based optical/NIR and millimeter telescopes allowed us to follow the afterglow evolution from about a minute from the high-energy event  down to the host galaxy limit. We discuss the physical parameters required to model these emissions.}
{The prompt emission of \object{GRB\,060908} was characterized by two main periods of activity, spaced by a few seconds of low intensity, with a tight correlation between activity and spectral hardness. Observations of the afterglow began less than one minute after the high-energy event, when it was already in a decaying phase, and it was characterized by a rather flat optical/NIR spectrum which can be interpreted as due to a hard energy-distribution of the emitting electrons. On the other hand, the X-ray spectrum of the afterglow could be fit by a rather soft electron distribution.}
{\object{GRB\,060908} is a good example of a gamma-ray burst with a rich multi-wavelength set of observations. The availability of this dataset, built thanks to the joint efforts of many different teams, allowed us to carry out stringent tests for various interpretative scenarios showing that a satisfactorily modeling of this event is challenging. In the future, similar efforts will enable us to obtain optical/NIR coverage comparable in quality and quantity to the X-ray data for more events, therefore opening new avenues to progress gamma-ray burst research.}


\maketitle

\section{Introduction}

The afterglows of gamma-ray bursts (GRBs) have attracted theoretical and observational interest. The difficulties in building a detailed and consistent model are indeed remarkable. In the context of the ``fireball" model, the blastwave is decelerated after sweeping up circumburst medium, and eventually enters a self-similar deceleration regime \citep{BlMc76}. The onset of the afterglow, when the fireball begins to decelerate, requires accurate relativistic computations in order to derive reliable (i.e. not only qualitative) predictions \citep[see e.g. ][]{BiRu05,KoZh07}. The scenario emerging from observations both in the X-rays and at longer wavelengths (optical, near-infrared; hereafter NIR) appears to be even more complicated than expected only a few years ago, with the superposition of emission from different mechanisms beyond the external shock emission. 

\textit{Swift}-XRT observations have provided an increasing sample of well-monitored observations, allowing for the derivation of the well-known ``canonical" X-ray light-curve \citep{Nou06}. As shown by many authors \citep[e.g. ][]{Pan06,Zha07b,Pan07,Tak07,Will10}, the early X-ray afterglow is often dominated by a steeply decaying emission component usually attributed to large-angle emission produced during the main burst \citep{Fen96,Tagl05,Zha06,Lia07}, although it has also been attributed to progressively fading central engine activity \citep{FaWe05,Fan08,Kum08}. The subsequent ``shallow-decay" phase \citep[for an overview, see][]{Lia07} is still attributed to prolonged central engine activity \citep{FaPi06,Joh06}, although there are other proposed mechanisms which can be at work: ejecta with a wide Lorentz $\Gamma$-factor energy distribution \citep{ReMe98}, varying micro-physics parameters \citep{Pan07}, delayed burst emission due to dust-scattering \citep[e.g.][who reach different conclusions]{ShDa07,She09}, off-axis initial observation \citep{Gran05,Donag06,Guid09}, etc. 

In the optical/NIR the situation is somewhat less defined. Robotic telescopes throughout the world have provided early-time light curves often fully overlapped in time to the \textit{Swift}-XRT (and -UVOT) observations. The data quality, however, is not always adequate for a detailed modelling, due to the modest aperture of most robotic, rapid-pointing, telescopes. For several events it was possible to detect the afterglow (external shock) onset \citep{Vestr06,Mol07,Fer09,Ryk09,Klo09} as predicted by semi-analytical estimates \citep{SaPi99} and more accurate numerical analyses \citep{KoZh07,JiFa07}. The lack of reverse shock \citep[see e.g.][]{Mund07,JiFa07} confirms the general results for \textit{Swift} GRBs obtained by the UVOT \citep{Rom06}. These (lack of) findings impose severe constraints on the micro-physics parameters of the relativistic shocks or alternatively suggests that additional ingredients, such as magnetically dominated outflows, are required \citep{Lyu03,Fan04,ZaKo05}. The prompt emission from \object{GRB\,990123} \citep{Ake99}, considered to be a typical example of reverse shock emission peaking in the optical, was also interpreted as the long wavelength tail of the large-angle high-energy emission from the prompt event \citep{PaKu07}. Reverse shock emission was invoked to model the early-time post-flash optical emission of the exceptionally bright \object{GRB\,080319B} \citep{Bloo09,KuPa08,Rac08,Yu09}. A comparable emission due to the reverse and forward shock was also proposed for \object{GRB\,070802} \citep{Kru08}. Partly motivated by the increasing difficulties of the so-called ``standard model" in interpreting the rich multi-wavelength datasets now available, there is a rising interest in alternative scenarios, such as the ``cannonball" model \citep{DaDa09,Dado09}. This scenario has shown remarkable fitting capabilities \citep[e.g.][]{DaDa10a,DaDa10} although it is still lacking of a comprehensive independent analysis campaign.

Some classification schemes have been proposed to interpret the rich variety displayed by optical afterglows. \citet{Zha03} and \citet{JiFa07} define three classes, depending on the mutual importance of the reverse and forward shock emission based on theoretical considerations. Class I is constituted by afterglows showing both the reverse and forward shock emissions, class II is for afterglows with a prominent reverse shock emission outshining all other components, and class III contains events characterised by forward shock emission only. 
\cite{PaVe08} observationally classify the optical afterglows in four classes following the temporal behaviour of the early optical emission and try to find a common scenario producing all the different observed behaviours: fast-rising with an early peak, slow-rising with a late peak, flat plateau, and rapid decays since the first measurements. They conclude that an emission due to the forward or reverse shock coming from a structured collimated outflow can explain all the four shapes, by varying the observer location and the structure of the outflows. However, the afterglows with plateaus and slow rises could also be due to a long-lived injection of energy in the blast wave. Interestingly, these authors find a possible peak flux-peak time correlation for the fast- (extended to slow-) rising optical afterglows that could provide a way to use them as standard candles. Note however that \citet{Klo09} and \citet{Kann09} with more data later questioned the tightness of the correlation. \citet{Lia09} discovered a set of correlations between afterglow onset parameters in the optical and GRB parameters, in particular a tight correlation between the initial Lorentz factor and the burst isotropic energy \citep[see also][]{DaDa10}.   
Moreover, as clearly demonstrated by the case of \object{GRB\,050820A} \citep{Vestr06} or \object{GRB\,080319B} \citep{Rac08}, the possible influence of the optical emission coming from the prompt GRB phase should also be taken into account when analysing the early light curve, further complicating the picture. 

In general, a satisfactory understanding of the early afterglow phases is still lacking. Events with high-quality optical early-time observations carried out with robotic telescopes and/or the UVOT are thus especially important for testing the predictions of different models. We discuss here the case of \object{GRB\,060908}. In Sect.\,\ref{sec:grb} we report the main observational data available for this event. In Sect.\,\ref{sec:data} we give some details of the data analysis for \textit{Swift}-BAT, -XRT, -UVOT, REM, SMARTS, Danish 1.54m, NOT, UKIRT, TNG and the Plateau de Bure Interferometer observations and in Sect.\,\ref{sec:disc} we discuss our results. Our main conclusions are given in Sect.\,\ref{sec:concl}.

\section{\object{GRB\,060908}}
\label{sec:grb}

\object{GRB\,060908} was detected by the \textit{Swift} satellite \citep{Geh04} on Sept. 8, 2006 at 08:57:22 UT \citep{Eva06}. Further analysis (see Sect.\,\ref{sec:databat}) yielded a revision of the GRB time. The optical afterglow was detected from ground by the PROMPT\footnote{http://www.physics.unc.edu/$\sim$reichart/prompt.html} telescope showing a bright ($r \sim 15$\,mag about 105\,s after the burst), rapidly fading source \citep{Nys06}. The optical afterglow was then confirmed at coordinates RA=02:07:18.3 and DEC=+00:20:31 (J2000, 0.5\arcsec\ error) with the REM telescope\footnote{http://www.rem.inaf.it} by \citet{Ant06} reporting $R \sim 17$\,mag about 7\,min after the burst. A first estimate of the decay rate was provided by \citet{Wie06} as $\alpha = 1.07 \pm 0.11$, with observations carried out with the Danish 1.54m telescope at ESO-La Silla equipped with DFOSC\footnote{http://www.ls.eso.org/lasilla/Telescopes/2p2T/D1p5M/}. Later observations were also reported by \citet{And06}.

\citet{Rol06} derived a redshift identifying the absorption lines of \ion{C}{IV} and \ion{Si}{II}, and possibly \ion{Al}{III} by means of observations performed with the Gemini-North telescope equipped with GMOS\footnote{http://www.gemini.edu/sciops/instruments/gmos/gmosIndex.html}. Their redshift estimate was later corrected by \citet{Fyn09} to $z = 1.884 \pm 0.003$.
The afterglow was not detected at 8.46\,GHz with the VLA\footnote{http://www.vla.nrao.edu/} a day after the burst with a $3\sigma$ upper limit of 77\,$\mu$Jy \citep{Cha06}. The effect of the host galaxy on the light-curve was initially detected by \citet{Tho06} with the NOT equipped with ALFOSC\footnote{http://www.not.iac.es/instruments/alfosc/}. 

Throughout the paper, the decay and energy spectral indices $\alpha$ and $\beta$ are defined by $F_\nu(t,\nu) \propto (t-T_0)^{-\alpha}\nu^{-\beta}$, where $T_0$ is the onset time of the burst. We assume a $\Lambda{\rm CDM}$ cosmology with $\Omega_{\rm m} = 0.27$, $\Omega_\Lambda = 0.73$ and $h_0 = 0.71$. At the redshift of the GRB ($z = 1.88$), the luminosity distance is $\sim 15$\,Gpc ($\sim 4.5 \times 10^{28}$\,cm, corresponding to a distance modulus $\mu = 45.8$\,mag). The Galactic extinction in the direction of the afterglow is E$_{B-V} = 0.03$\,mag \citep{Sch98}. All errors are $1\sigma$ unless stated otherwise.

\section{Observations, data reduction and analysis}
\label{sec:data}

\subsection{\textit{Swift}-BAT}
\label{sec:databat}

\begin{figure}
\includegraphics[width=\columnwidth]{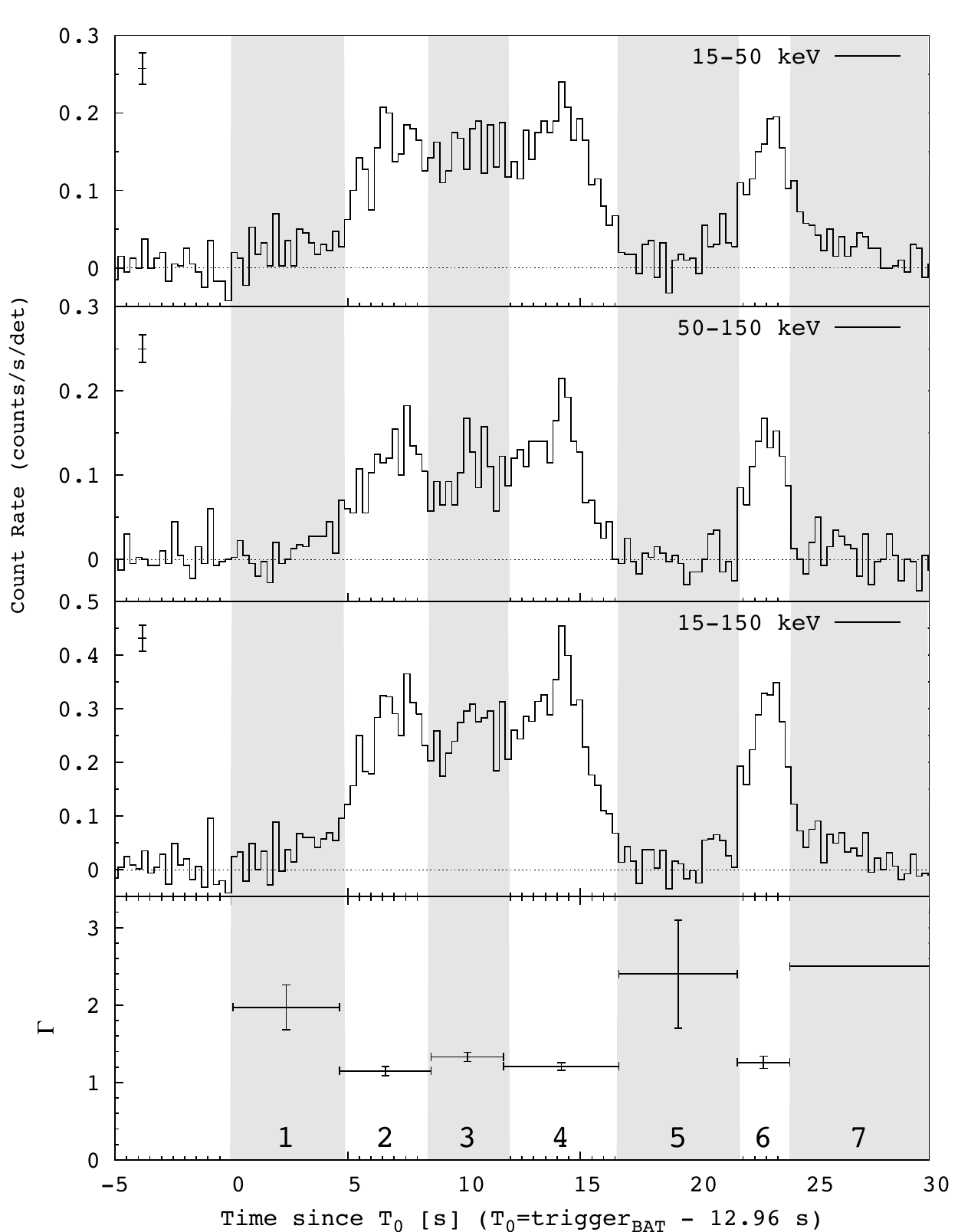}
\caption{BAT mask-weighted light curves of the prompt emission in the 15--50, 50--150 and 15--150\,keV energy bands from the top downward, respectively. Times refer to the revised BAT trigger time ($T_0=T_{\rm BAT}-12.96$\,s) The vertical bar shown at the top left corner in each plot shows the typical error on the count rate. The time binning is 0.256\,s. The count rates are expressed in units of counts\,s$^{-1}$ per fully illuminated detector for an equivalent on-axis source. The bottom panel shows the spectral photon index as a function of time.}
\label{fig:stacked}
\end{figure}

\begin{figure}
\includegraphics[width=\columnwidth]{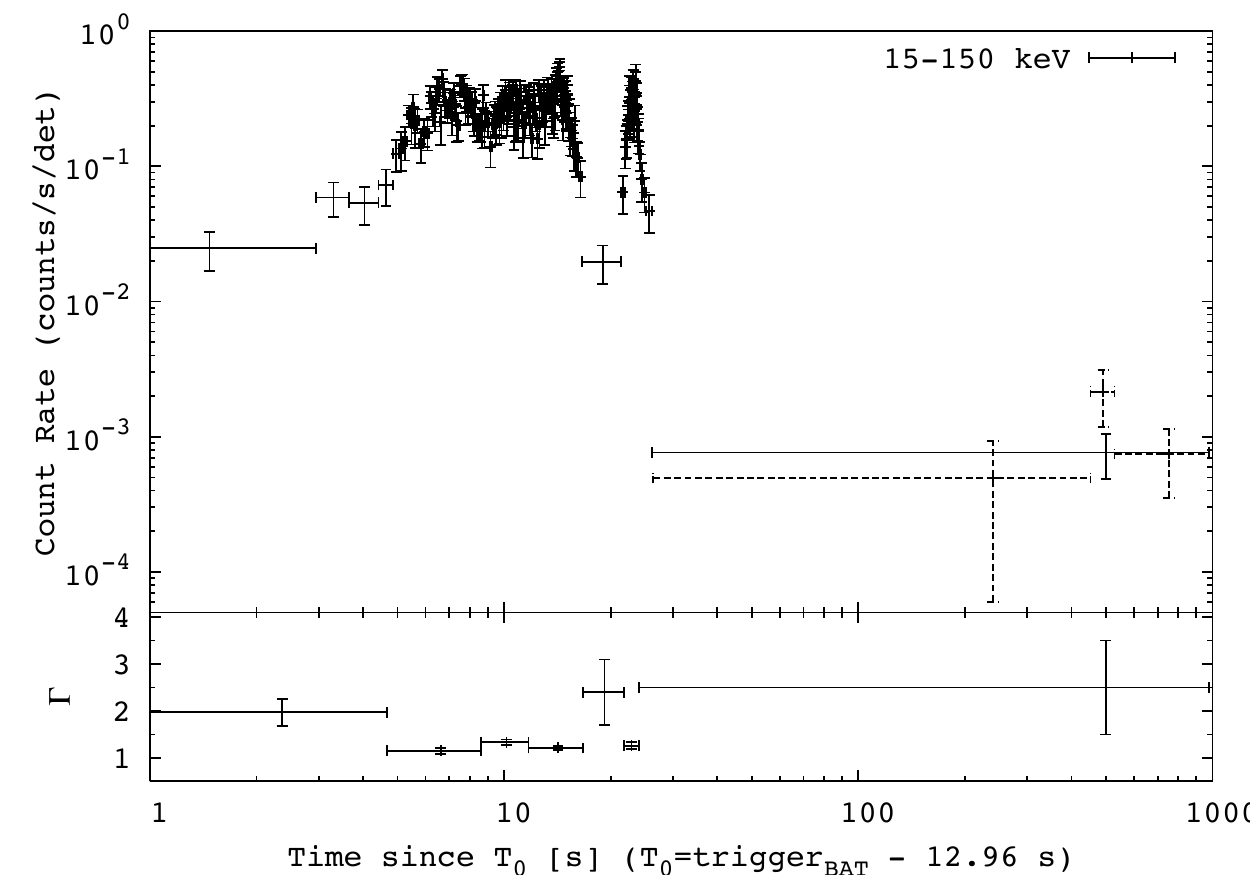}
\caption{Top panel: 15--150\,keV mask-weighted light curve of the prompt emission up to 1000\,s after the trigger time. The binning is variable and was determined so as to have at least 3$\sigma$ significant rates in each bin, except for the last point from $T_0+26.1$\,s to $T_0+975.2$\,s which is at 2.7$\sigma$ ($T_0=T_{\rm BAT}-12.96$\,s). Dashed lines show the last point split into three bins: in particular we note detection of emission from $T_0+453$\,s to $T_0+529$\,s the 2.2$\sigma$. Bottom panel: temporal evolution of the spectral photon index $\Gamma$.}
\label{fig:all}
\end{figure}

\object{GRB\,060908} triggered BAT at 08:57:22.34\,UT, which hereafter will be referred to as $T_{\rm BAT}$. We extracted the mask-weighted light curves and energy spectra in the 15--150\,keV band following the BAT team instructions\footnote{http://heasarc.nasa.gov/docs/swift/analysis/threads/bat\_threads.html}. The 15--150\,keV prompt emission profile consists of an initial structure where three pulses can be identified lasting about 15\,s, followed by a 5\,s long quiescent time ended by another isolated pulse comparable with the previous ones (Fig.~\ref{fig:stacked}). The total duration (15--150\,keV) in terms of $T_{90}$ is $19.3\pm0.3$\,s \citep{Pal06}. We note that the onset of the GRB, $T_0$, occurs before the trigger time $T_{\rm BAT}$: in particular, we find $T_0=T_{\rm BAT}-12.96$\,s from significance requirements described below. This was also pointed out by the BAT team \citep{Pal06}. Also worth mentioning is the evidence of a weak prolonged emission at high energies, when the light curves were binned to reach a given significance in the count rate of each bin. Fig.\,\ref{fig:all} shows the case of the 15--150\,keV profile when 3$\sigma$ are required for each time bin. The last point from $T_0+26.1$\,s to $T_0+975.2$\,s is 2.7$\sigma$ significant. Notably, this is broadly concurrent with some flaring activity in the soft X-rays detected by the {\em Swift}/XRT (see Sect.\,\ref{sec:dataxrt}). 

Energy spectra were extracted in seven contiguous time intervals as reported in Fig.\,\ref{fig:stacked}. The choice was driven by the light curve evolution:
we identified the first rise (1), the three overlapping pulses of the first structure (2--4), the quiescent time (5), the following isolated pulse (6) and the final long weak tail (7). All of the spectra can be fit with a single power law. Detailed results of the spectral fitting are reported in Table\,\ref{tab:BAT}.

\begin{table}
\begin{tiny}
\caption{BAT energy spectra (15--150\,keV) in the seven distinct time intervals of the prompt emission. Each interval is fit with a power law ($\Gamma$ is the photon index). Times are referred to the revised BAT trigger time ($T_0=T_{\rm BAT}-12.96$\,s). Uncertainties are 1$\sigma$.\label{tab:BAT}}
\centering
\begin{tabular}{llllll} \hline \hline
\# & $T_{\rm start}$ &  $T_{\rm stop}$ &  $\Gamma$     & Fluence & $\chi^2$/dof \\
   & (s)             &   (s)           &               & ($10^{-7}$~erg~cm$^{-2}$) &       \\\hline 
1  & $0.00$        & $4.66$          & $1.97\pm0.29$ &    $1.0\pm0.2$     &  $11.63/7$  \\
2  & $4.66$          & $8.59$         & $1.15\pm0.06$ &    $6.8\pm0.25$    & $47.44/47$  \\
3  & $8.59$         & $11.71$         & $1.33\pm0.06$ &    $5.5\pm0.22$    & $41.20/41$   \\
4  & $11.71$         & $16.66$           & $1.21\pm0.05$ &    $9.0\pm0.3$     & $35.80/47$  \\
5  & $16.66$           & $21.76$           & $2.4\pm0.7$   &    $0.54\pm0.19$   & $0.05/1$    \\
6  & $21.76$           & $24.00$         & $1.26\pm0.08$ &    $3.8\pm0.2$     & $46.43/43$  \\
7  & $24.00$         & $975.3$         & $2.5\pm1.0$   &    $2.8\pm1.5$     & $3.33/4$   \\
total & $0.0$     & $28.0$          & $1.36\pm0.04$ &    $27.6\pm0.6$    & $46.8/46$   \\
\hline
\end{tabular}
\end{tiny}
\end{table}

Figures\,\ref{fig:stacked} and \ref{fig:all}  display the 15--150\,keV light curve and the evolution of the spectral photon index. We can see a hardening around the peaks of the pulses similar to the canonical behaviour of other GRBs. Apart from this, the prompt emission does not exhibit strong spectral evolution, the photon index being pretty constant along different pulses and consistent with that derived from the total spectrum, $\Gamma_{\rm tot}=1.36\pm0.04$. The total fluence in the 15--150\,keV band is $(2.8\pm0.1)\times10^{-6}$\,erg\,cm$^{-2}$, $\sim$10\% of which from the weak long tail after $T_0+24.00$\,s. These results are also in agreement with those by \citet{Pal06}.

Unfortunately, no measurement of the peak energy $E_{\rm p}$ of the time-integrated spectrum is available. However, from the hardness of $\Gamma_{\rm tot}$ we can infer that $E_{\rm p}$ lies above the BAT passband or close to its upper bound.
Indeed, if we fit the total spectrum with a Band function \citep{Band93} and fix the high energy photon index $\beta$ at the typical value $\beta= -2.3$, we find for the low-energy index $\alpha = -0.9\pm0.3$ and $E_{\rm p}=133_{-33}^{+120}$\,keV with $\chi^2/{\rm dof} = 38.5/45$ (90\% errors), in fair agreement with the estimate by \citet{Ghi08} and \citet{Sak09}. This is also consistent with the empirical relation between $E_{\rm p}$ and $\Gamma$ (measured by BAT) for a number of bursts \citep{Zha07}. Assuming this value for $E_{\rm p}$, we derive the following values: $E_{\rm p,i}=383_{-95}^{+346}$\,keV (rest-frame peak energy) and $E_{\rm iso}=(6.2\pm0.7)\times10^{52}$\,erg (isotropic-equivalent released energy in the rest-frame $1-10^{4}$\,keV energy band, errors at 90\%).

\subsection{\textit{Swift}-XRT}
\label{sec:dataxrt}

The XRT observations of \object{GRB\,060908} started at 08:58:42\,UT, about $\sim 80$ s after the BAT trigger, and ended on September 20, 2006 at 23:01:56\,UT. The XRT afterglow candidate alert was delivered about 100\,s after the BAT trigger. The monitoring consisted of 14 different observations. The first data were taken in windowed timing (WT) mode and lasted for $\sim 100$\,s. After that, the on-board measured count rate was low enough for the XRT to switch to the photon counting (PC) mode; for the rest of the follow-up, XRT remained in PC mode. 

The XRT data were reduced using the {\em xrtpipeline} task (v.2.5), applying standard calibration and filtering criteria, i.e., we cut out temporal intervals in which the CCD temperature was above $-47\,^\circ$C and we remove hot and flickering pixels. An on-board event threshold of $\sim 0.2$\,keV was applied to the central pixel; this was proven to reduce most of the background due to the bright Earth and/or the CCD dark current. We selected XRT grades 0-2 and 0-12 for WT and PC data, respectively.

The intensity of the source was high enough to cause significant pile-up in the first part of the PC mode observations. In order to correct for the pile-up, we extracted the counts from an annulus with an inner radius of 4 pixels and an outer radius of 20 pixels (9\arcsec\ and 47\arcsec, respectively). We then corrected the observed count rate for the fraction of the XRT point spread function (PSF) lying outside the extraction region. Data in WT mode were not affected by pileup; thus, for WT observations and for the remaining PC observations, a region of 20 pixel radius was selected. Physical ancillary response files were generated using the task {\em xrtmkarf}, to account for different extraction regions.

\begin{table}
\caption{Spectral analysis of the \textit{Swift}-XRT data. The switch from WT to PC mode occurred at $\sim 180$\,s after the burst. Analysis of data later than about 2000\,s gave results comparable to those for the PC mode but with lower statistical significance.\label{tab:XRTspec}} 
\centering
\begin{tabular}{rcccc} 
\hline 
\hline
Mode & N$_H$                            & N$_{\rm H}$     & Photon index & $\chi^2$/dof \\
	&					& (rest frame) \\
           &   ($10^{21}$cm$^{-2}$)      & ($10^{21}$cm$^{-2}$)     &                         & \\
\hline
WT     & $7.2_{-4.6}^{+5.2}$ & - & $2.32_{-0.29}^{+0.33}$  & 17.9/17 = 1.05 \\
PC     & $14.1_{-6.0}^{+7.0}$ & - & $2.32_{-0.20}^{+0.20}$  & 11.9/17 = 0.70 \\
\hline
WT     & $0.23$ & $3.6_{-3.0}^{+3.6}$ & $2.28_{-0.17}^{+0.20}$  & 17.7/17 = 1.03 \\
PC     & $0.23$ & $8.3_{-3.7}^{+5.7}$ & $2.17_{-0.22}^{+0.25}$  & 12.9/17 = 0.76 \\
\hline
\end{tabular}
\end{table}

For spectral analysis we used redistribution matrices version\,11. Spectral fit results are reported in Table\,\ref{tab:XRTspec}. The spectra were modeled with a simple absorbed power-law. The Galactic column density around the GRB\,060908 position is $2.3\times 10^{20}$ cm$^{-2}$ \citep{Kal05}. By fitting a power law model with a Galactic contribution fixed at the above value, we derive an intrinsic column density of $8.3^{+5.7}_{-3.7}\times 10^{21}$ cm$^{-2}$ for data
collected in PC mode, where spectral variability is lower. This is in line with rest-frame absorption observed in GRBs \citep{Camp10}.

\begin{figure*}
\includegraphics[width=\textwidth]{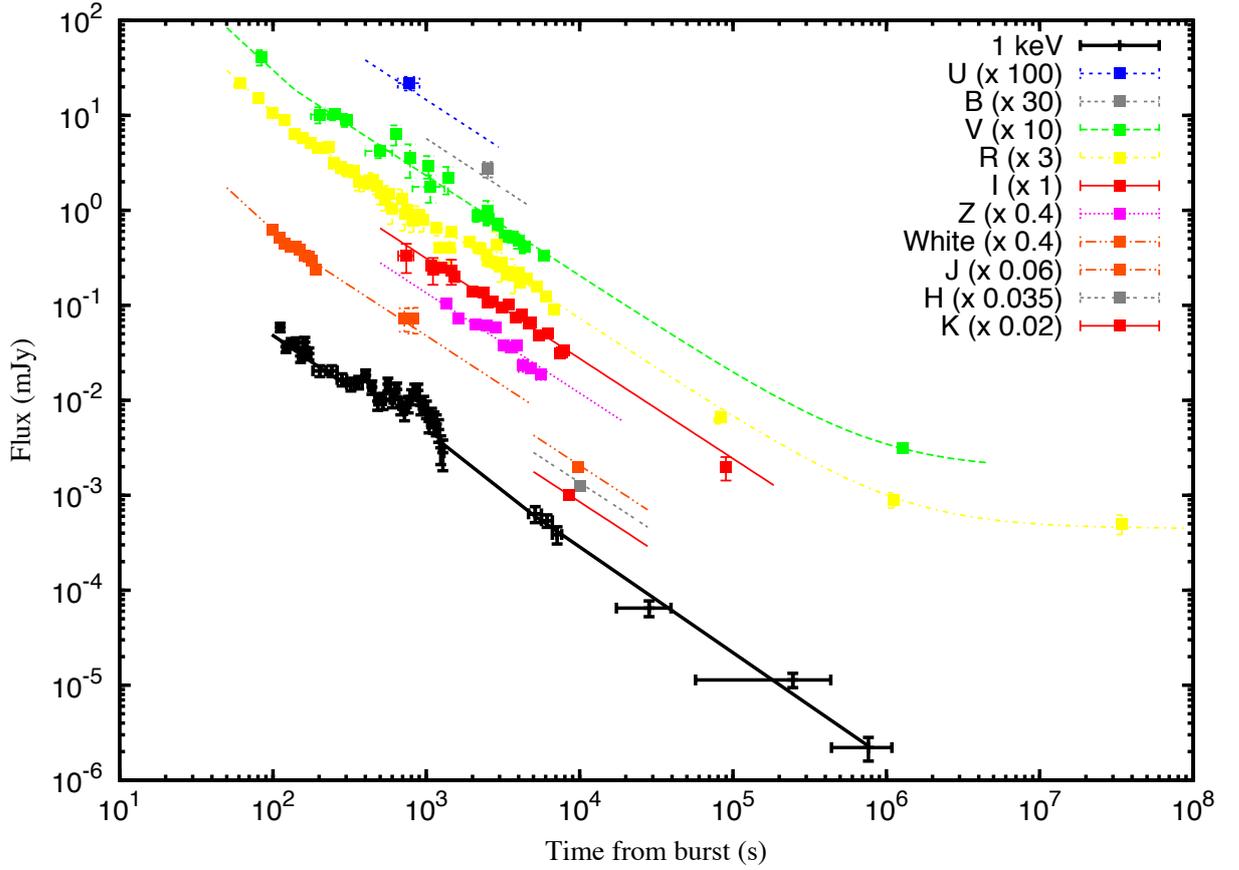}
\caption{Light curve in the X-ray and optical/NIR bands for the afterglow of \object{GRB\,060908}. The light curves are fitted with a simple power-law with index $\sim 1.1$ for the X-ray data while two power-laws smoothly joined are applied to optical/NIR data. The time delay from the burst was corrected as $T_0 = T_{\rm BAT}-12.96$\,s (see Sect.\,\ref{sec:databat}). }
\label{fig:lctot}
\end{figure*}

\begin{figure}
\includegraphics[width=\columnwidth]{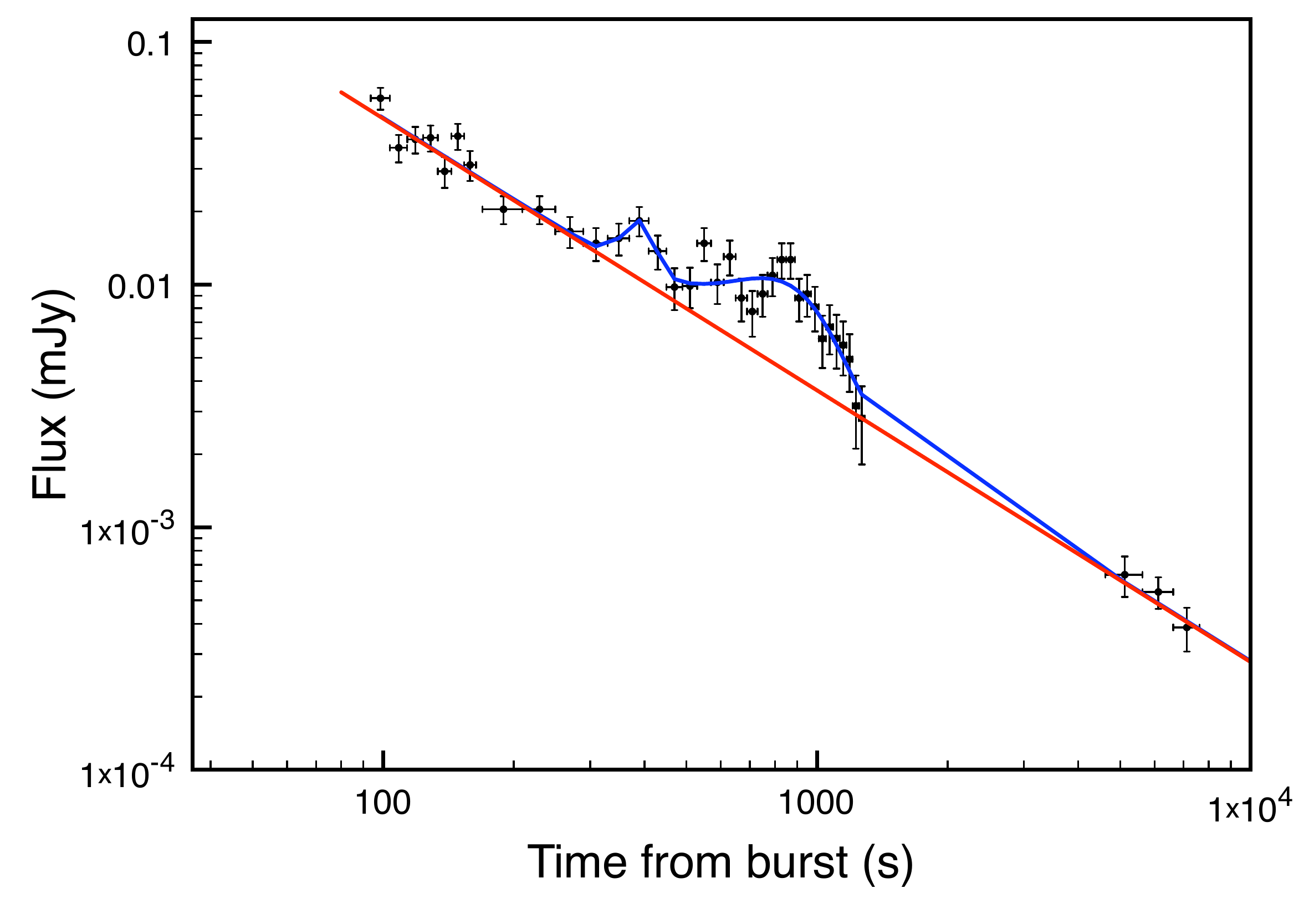}
\caption{Light curve at 1\,keV for the \textit{Swift}-XRT observation around the epoch of the small flares superposed on the afterglow power-law decay. The blue  solid line is a fit with a temporal power-law decay with index $\sim 1.1$ and two Gaussians centred at $\sim 390$ and $\sim 810$\,s from the burst. The red line shows the power-law only component. The time delay from the burst was corrected as $T_0 = T_{\rm BAT}-12.96$\,s (see Sect.\,\ref{sec:databat}). }
\label{fig:lcxrt}
\end{figure}

The total light curve in physical units is shown in Fig.\,\ref{fig:lctot}. The curve is characterised by a constant power-law decay with index $\alpha \sim 1.1$, while from about 200 to 1000\,s a complex, although not strongly dominant, flaring activity is superposed on the continuous decay. Apart from these flares, the decay goes on uninterrupted up to the last XRT observations at $\sim 10^{6}$\,s from the burst. We could model ($\chi^2/{\rm dof} =  35.6/33 = 1.08$) the XRT light curve with a simple power-law with decay index $\alpha_{\rm X} = 1.12_{-0.02}^{+0.05}$, plus two Gaussians (Fig.\,\ref{fig:lcxrt} to fit the main flares. Since we are mainly interested in the behaviour of the underlying afterglow, the Gaussian function representing the flares was chosen for simplicity, and no physical meaning is attributed to them. A more detailed analysis of flaring activity in this and other events is reported in \citet{Chi10}. 
\citet{Lia08} claimed a possible identification of a break in the \textit{Swift}-XRT light-curve. \object{GRB\,060908} was indeed classified as part of their ``bronze'' class, i.e. events showing a break with post-break decay index steeper than 1.5. This is due to the neglecting of the effect of the end of the flaring activity at about 1000\,s.

\subsection{Optical/NIR data}
\label{sec:optnir}

\textit{Swift}-UVOT \citep{Rom05} data were retrieved from the HEASARC public archive\footnote{http://heasarc.gsfc.nasa.gov/cgi-bin/W3Browse/swift.pl}. UVOT data analysis was carried out following standard recipes\footnote{http://heasarc.gsfc.nasa.gov/docs/swift/analysis/}. The data were screened for standard bad times, South Atlantic Anomaly passages, Earth limb avoidance, etc. The task {\em uvotsource} was applied to compute aperture photometry for images, and {\em uvotevtlc} for event files. Photometry in the $UBV$ broad band filters and without filter (``white") was derived with radii of 6\,\arcsec\   and 12\,\arcsec\  of aperture for image and event file analysis, respectively. For the bluer filters UVW1, UVM2 and UVW2, a radius twice as large was used. We also verified the consistency between image and event file photometry for a set of bright, isolated, unsaturated stars. Consistency between UVOT and REM photometry for the $V$ band was also checked.
The UVOT alert was delivered about 15\,min after the BAT trigger but  UVOT observations had already started about one minute after the trigger. The results are reported in Table\,\ref{tab:totlc}.

\begin{table}
\begin{tiny}
\caption{Light-curve data of the \object{GRB\,060908} afterglow obtained by our collaboration. The time delay from the burst was updated as $T_0 = T_{\rm BAT}-12.96$\,s (see Sect.\,\ref{sec:databat}). Galactic extinction has not been removed from these data. \label{tab:totlc}} 
\centering
\begin{tabular}{crrcc} 
\hline 
\hline
Band & $T-T_0$ & Bin half size & Magnitude & Telescope\\
              &  (s)                    &  (s) &  \\
\hline 
$U$ & 778 & 125 & $17.05 \pm 0.18$ & UVOT \\
\hline
$B$ & 2522 & 22.5 & $19.10 \pm 0.20$ & SMARTS \\
\hline
$V$ & 84 & 2.5 & $14.87 \pm 0.20$ & UVOT \\ 
& 202 & 25 & $16.38 \pm 0.21$ & UVOT \\ 
& 252 & 25 & $16.37 \pm 0.15$ & UVOT \\ 
& 302 & 25 & $16.52 \pm 0.17$ & UVOT \\ 
& 500 & 100 & $17.34 \pm 0.17$ & UVOT \\ 
& 633 & 10 & $16.88 \pm 0.26$ & REM \\
& 783 & 10 & $17.50 \pm 0.42$ & REM \\
& 1034 & 30 & $17.71 \pm 0.29$ & REM \\
& 1063 & 250 & $18.28 \pm 0.35$ & UVOT \\ 
& 1385 & 30 & $18.04 \pm 0.35$ & REM \\
& 2522 & 15 & $18.90 \pm 0.30$ & SMARTS \\
& 1278472 & 2100  & $25.04 \pm 0.10$ & NOT \\
\hline
$R$ & 61 & 5 & $14.02 \pm 0.04$ & REM \\
& 80 & 5 & $14.43 \pm 0.06$ & REM  \\
& 100 & 5 & $14.83 \pm 0.06$ & REM  \\
& 119 & 5 & $15.00 \pm 0.08$ & REM  \\
& 138 & 5 & $15.38 \pm 0.10$ & REM  \\
& 157 & 5 & $15.48 \pm 0.12$ & REM  \\
& 176 & 5 & $15.60 \pm 0.12$ & REM  \\
& 195 & 5 & $15.73 \pm 0.12$ & REM  \\
& 214 & 5 & $15.74 \pm 0.13$ & REM  \\
& 233 & 5 & $15.71 \pm 0.14$ & REM  \\
& 248 & 10 & $16.14 \pm 0.14$ & REM  \\
& 277 & 10 & $16.25 \pm 0.15$ & REM  \\
& 306 & 10 & $16.32 \pm 0.16$ & REM  \\
& 336 & 10 & $16.35 \pm 0.16$ & REM  \\
& 365 & 10 & $16.63 \pm 0.21$ & REM  \\
& 394 & 10 & $16.67 \pm 0.20$ & REM  \\
& 423 & 10 & $16.63 \pm 0.18$ & REM  \\
& 452 & 10 & $16.58 \pm 0.20$ & REM  \\
& 481 & 10 & $16.73 \pm 0.21$ & REM  \\
& 510 & 10 & $16.93 \pm 0.26$ & REM  \\
& 542 & 10 & $17.11 \pm 0.25$ & REM  \\
& 571 & 10 & $16.95 \pm 0.23$ & REM  \\
& 601 & 10 & $17.34 \pm 0.34$ & REM  \\
& 695 & 10 & $17.08 \pm 0.28$ & REM  \\
& 724 & 10 & $17.48 \pm 0.36$ & REM  \\
& 753 & 10 & $17.36 \pm 0.33$ & REM  \\
& 825 & 30 & $17.65 \pm 0.27$ & REM  \\
& 894 & 30 & $17.49 \pm 0.24$ & REM  \\
& 964 & 30 & $17.64 \pm 0.27$ & REM \\
& 1211 & 65 & $18.36 \pm 0.36$ & REM  \\
& 1420 & 136 & $18.37 \pm 0.34$ & REM  \\
& 2177 & 30 & $18.36 \pm 0.07$ & Danish \\ 
& 2461 & 30 & $18.56 \pm 0.06$ & Danish \\ 
& 2522 & 15 & $18.70 \pm 0.10$ & SMARTS \\
& 2652 & 60 & $18.72 \pm 0.06$ & Danish  \\ 
& 2864 & 25 & $18.29 \pm 0.40$ & REM  \\
& 2874 & 60 & $18.79 \pm 0.06$ & Danish \\ 
& 3100 & 201 & $18.77 \pm 0.40$ & REM  \\
& 3125 & 90 & $18.86 \pm 0.06$ & Danish \\ 
& 3402 & 90 & $19.00 \pm 0.06$ & Danish \\ 
& 3683 & 90 & $19.06 \pm 0.06$ & Danish \\ 
& 3787 & 334 & $19.03 \pm 0.43$ & REM  \\
& 4018 & 120 & $19.02 \pm 0.05$ & Danish \\ 
& 82747 & 7317 & $22.83 \pm 0.16$ & Danish \\ 
& 1113287 & 2400 & $25.00 \pm 0.20$ & NOT \\
& 34343882 & 2865 & $25.63 \pm 0.25$ & TNG \\
\hline
$I$ & 740 & 85 & $17.17 \pm 0.37$  & REM   \\
& 1105 & 30 & $17.52 \pm 0.34$ & REM   \\
& 1456 & 30 & $17.55 \pm 0.32$ & REM   \\
& 2522 & 22.5 & $18.40 \pm 0.10$ & SMARTS \\
\hline
$J$ & 9819 & 405 & $19.22 \pm 0.06$  & UKIRT \\
\hline
$H$ & 10039 & 405 & $18.65 \pm 0.08$ & UKIRT  \\
\hline
$K$ & 8586 & 520 & $17.80 \pm 0.03$ & UKIRT \\
\hline
White & 100 & 5 & $15.23 \pm 0.11$ & UVOT \\ 
& 110 & 5 & $15.44 \pm 0.10$ & UVOT \\ 
& 120 & 5 & $15.60 \pm 0.10$ & UVOT \\ 
& 130 & 5 & $15.67 \pm 0.10$ & UVOT \\ 
& 140 & 5 & $15.69 \pm 0.10$ & UVOT \\ 
& 150 & 5 & $15.76 \pm 0.10$ & UVOT \\ 
& 160 & 5 & $15.92 \pm 0.11$ & UVOT \\ 
& 170 & 5 & $15.94 \pm 0.10$ & UVOT \\ 
& 180 & 5 & $16.06 \pm 0.11$ & UVOT \\ 
& 190 & 5 & $16.28 \pm 0.12$ & UVOT \\ 
& 720 & 25 & $17.57 \pm 0.30$ & UVOT \\ 
& 820 & 25 & $17.58 \pm 0.33$ & UVOT \\ 
\hline
\end{tabular}
\end{tiny}
\end{table}

REM is a 60\,cm diameter fast-reacting (10$^\circ$\,s$^{-1}$ pointing speed) telescope located at the Cerro La Silla premises of the European Southern Observatory (ESO), Chile \citep{Zer01,Chi03,Cov04a,Cov04b}. The telescope hosts REMIR, an infrared imaging camera, and ROSS, an optical imager and slitless spectrograph. The two cameras observe simultaneously the same field of view of $10$\arcmin$\times10$\arcmin\ thanks to a dichroic. Unfortunately, REMIR could not observe this GRB due to maintenance work. The \textit{Swift}-BAT alert was received by the REM telescope 14.7\,s after the BAT trigger time. The telescope reacted automatically and was tracking the GRB field 34.1\,s after receipt of the alert (48.8\,s after the BAT trigger). ROSS data ($V, R$ and $I$ bands) and other ground based telescope data were reduced in a standard way by means of tools provided by the ESO-Eclipse package \citep{Dev97}. Photometry for REM data and other ground based telescopes was carried out with SExtractor \citep[v. 2.5.0;][]{BeAr95}. Photometric calibration was accomplished by using instrumental zero points, checked with observations of standard stars in the \object{SA\,110} Landolt field (Landolt 1992). The results are reported in Table\,\ref{tab:totlc}.

We also obtained data for two nights using the Danish 1.54 m telescope at ESO - La Silla, Chile. We used the Danish Faint Object Spectrograph and Camera (DFOSC) instrument, which has a 13\arcmin.7 $\times$ 13\arcmin.7\  field of view, with a pixel scale of 0\farcs395 per pixel. Our observations on the first night consisted of a series of short exposures in the $R$ band (and one in the $I$ band), increasing in exposure time to obtain comparable photometric uncertainties in each datapoint. The data taken on the second night consist of a series of $R$ band images with exposure ranging from 5 to 10\,min. Multicolour observations within 1\,hr after the burst were also obtained with the SMARTS 1.3\,m telescope\footnote{http://www.astro.yale.edu/smarts/smarts1.3m.html}. Calibration was carried out by means of secondary standard stars derived from the calibration of the REM data. 

NIR observations about 10\,ks after the GRB have been provided by the UKIRT 3.8m telescope at Mauna Kea, Hawaii Islands, USA\footnote{http://www.jach.hawaii.edu/UKIRT/}. We used the WFCAM wide field camera ($0\degr.75 \times 0\degr.75$). The data were reduced and analysed following standard NIR recipes. We carried out late time observations with the Nordic Optical 2.5m Telescope equipped with the ALFOSC and the TNG\footnote{http://www.tng.iac.es/} equipped with DOLoRes. Both telescopes are located at the Canarian island of La Palma. These observations were aimed at detecting the host galaxy of \object{GRB\,060908}. NOT observations were carried out under variable meteorological conditions on 2006 Sept. 21 ($R$ band) and 23 ($V$ band), about 12-14\,days after the GRB). TNG observations were carried out under good observing conditions on 2007 Oct. 10, more than one year after the burst. An object compatible with the afterglow position was visible. In consideration of the long delay this might well be the host galaxy ($R \sim 25.6$, Table\,\ref{tab:totlc}). Reduction was performed in a standard way and calibration was carried out by using secondary standards in the field. Finally, we used published optical data obtained with the Palomar 60\,inch telescope from \citet{Cen09}.

\begin{figure}
\includegraphics[width=\columnwidth]{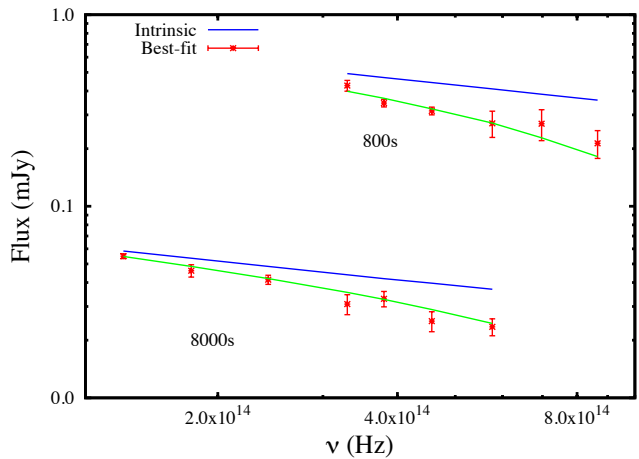}
\caption{Optical/NIR SEDs obtained at 800 and 8000s.  The best-fit model including Milky-Way and rest-frame absorption and the intrinsic spectral shape are reported.}
\label{fig:optsed}
\end{figure}

The total light curve is shown in Fig.\,\ref{fig:lctot} (see also Table\,\ref{tab:totlc}). Optical data were fitted by using power-law models for both the light-curve  and spectra. The light-curve appears characterised by an initial steeper decay \citep[see also][]{Oat09, Kann09} with $\alpha_{\rm optNIR} \sim 1.4$ followed by a flattening, $\alpha_{\rm optNIR} \sim 1$ with a transition time of about 120-360\,s. At later times the optical light curve began to be dominated by the host galaxy. Any late-time break is therefore difficult to locate. 
There was no clear spectral evolution. A  single rather blue power-law, $\beta_{\rm optNIR} \sim 0.3$, provided a satisfactory description (Fig.\,\ref{fig:optsed}) with some local, unconstrained, rest-frame absorption, E$_{B-V} \sim 0.03$, following the Small Magellanic Cloud (SMC) extinction curve \citep{Pei92}. Extinction curves typical of the Large Magellanic Cloud, the Milky Way \citep{Pei92} or starburst galaxies \citep{Calz00} environments gave worse fits. Forcing a late break improves the global fit significantly, though the break time is not well constrained. Considering that the number of different telescopes, filters, observing conditions, etc. might have introduced some inhomogeneity in the data, and artificially pushed the resulting $\chi^2$ up, we consider both possibilities (two or three power-laws in time) in the later discussion. Best-fit parameters and corresponding errors are reported in Table\,\ref{tab:optfit}.

\begin{table}
\caption{Optical/NIR light curve best fit results for models consisting of two (2PL) or three power laws (3PL). Spectral data are always fit with a single power-law plus possible rest-frame absorption following the SMC extinction curve and the known Galactic absorption. Upper limits are at 95\% confidence level. \label{tab:optfit}} 
\centering
\begin{tabular}{lcc} 
\hline 
\hline
 & 2PL & 3PL \\
\hline
$\alpha_1$ &  $1.48_{-0.25}^{+0.24}$ & $1.32_{-0.13}^{+0.19}$ \\
$t_{{\rm b}_1}$ (s) & $138_{-43}^{+167}$ & $365_{-228}^{+411}$ \\
$\alpha_2$ & $1.05_{-0.03}^{+0.03}$ & $0.94_{-0.36}^{+0.10}$ \\
$t_{{\rm b}_2}$ (s) & - & $2360_{-900}^{+4300}$ \\
$\alpha_3$ & - & $1.15_{-0.06}^{+0.24}$ \\
$\beta$ & $0.17_{-0.40}^{+0.34}$ & $0.33_{-0.29}^{+0.25}$ \\
E$_{B-V}$ (mag) & $< 0.10$ & $< 0.10$ \\
$\chi^2/{\rm dof}$ &  $159.1/121 = 1.32$ & $133.8/119 = 1.12$ \\
\hline
\hline
\end{tabular}
\end{table}

\subsection{Millimetre observations}
\label{sec:mm}
We complete our dataset with limits at millimetre wavelengths obtained with the Plateau de Bure Interferometer \citep{Guil92}. We observed the field at a mean time of 02:02 UT on 2009 Sept. 9 (17\,hr after the burst onset) simultaneously at 92 GHz (3 mm) and 236\,GHz (1\,mm) using the 5Dq compact five antenna configuration. Data calibration was done using the GILDAS software package\footnote{GILDAS is the software package distributed by the IRAM Grenoble GILDAS group.} using  MWC349 as flux calibrator, 3C454.3 as bandpass calibrator and 0235+164 as amplitude and phase calibrator. We did not detect any source at the position of the GRB afterglow with 3$\sigma$ limit of 1.17 mJy in the 92\,GHz band and 9.9\,mJy in the 236\,GHz band. This is consistent with the limit reported by \citet{Cha06} (see Sect.\,\ref{sec:grb}).

\section{Discussion}
\label{sec:disc}

\subsection{Prompt emission}
\label{sec:pe}

The prompt emission of \object{GRB\,060908} lasted $19.3 \pm 0.3$\,s \citep{Pal06}. The corresponding spectra (Table\,\ref{tab:BAT}) were typical of the long-soft class of GRBs \citep{Kouv93}, although the average photon index for the \object{GRB\,060908} prompt emission is rather close to the hard tail for the long/soft GRB distribution \citep{Sak08}. Spectral lag was also originally proposed by \citet{Nor00} and \citet{NorBo06} as a possible tool to better discriminate between GRB classes. Recently, \citet{Ukw09}, in a comprehensive study of spectral lags for a sample of 31 GRBs with measured redshift, reported for \object{GRB\,060908} a spectral lag consistent with zero. However, the errors were large enough to prevent firm conclusions. With the estimated isotropic and peak energies, \object{GRB\,060908} would lie within 2$\sigma$ of the ``Amati" relation \citep{Ama02, Ama06}. Applying the $E_{\rm iso}-\Gamma_0$ relation singled out by \citet{Lia09} the initial Lorentz factor should be $\Gamma_0 \sim 300$ (see Sect.\,\ref{sec:ea}).

No precursor was seen in the \textit{Swift}-BAT light-curve and the high energy emission did not show any detectable spectral evolution. The light-curve showed two periods of activity separated by a pause lasting a few seconds. The spectra during activity periods were substantially harder than that in the relatively quiescent interval; this is in agreement with previous findings about a general intensity-hardness correlation during prompt emission \citep{Gole83,Borg01}, which has been attributed to the curvature effect by \citet{Qin09}. After the prompt emission a longer-lasting soft emission is detectable, possibly up to about 1000\,s.

\subsection{The early afterglow}
\label{sec:ea}

The pulse at about 23\,s after the beginning of the prompt emission might mark the onset of the afterglow, which can usually be hidden by longer prompt activity. In this case the duration of the observed prompt emission would be just $\sim 15$\,s. Following this hypothesis, it is possible to estimate the initial Lorentz bulk motion $\Gamma_0$ by using the method describeed in \citet{Mol07}. The dependence of $\Gamma_0$ on environment parameters is weak and we can assume a constant density circumburst medium due to the rapid increase in flux before the onset. Under these assumptions in the so-called ``thin-shell" case, and applying Eq.\,1 in \citet{Mol07} where $\Gamma_0 = 2 \Gamma$, the initial Lorentz factor turns out to be $\Gamma_0 \sim 700\,(\eta_{0.2}n_0)^{-1/8}$, where $\eta$ in units of $0.2$ is the radiative efficiency and $n_0$ the circumburst constant number density in cm$^{-3}$.  This figure is in agreement with theoretical expectations \citep{ZhMe04} and close to recent estimates for a few GRBs in which prompt GeV photon emission was detection with the Fermi satellite \citep[e.g.][]{Abd09,Dep10,Ghi10}. This hypothesis, though intriguing, has also substantial difficulties. The last pulse of the prompt emission shows a time profile comparable to that of the other prompt emission pulses, suggesting a common origin. Moreover, it is characterized by a rather hard photon index, $\Gamma = 1.26 \pm 0.08$, comparable, as noted above, to the spectral parameters of the periods of high activity of the prompt emission. While similar to the optical/NIR spectral index, it is much harder than the initial X-ray afterglow spectral index (and of the long-lasting BAT emission detected at the end of the prompt phase). Such a complex spectral shape strongly suggests that the last pulse is part of the prompt emission, and is not related to the forward shock. In order to interpret the last pulse of the prompt emission as the afterglow onset we at least need to assume that the spectrum of the X-ray and optical/NIR afterglow observed about a minute after the high-energy event has already spectrally evolved remarkably soon after the onset.

Following the \citet{PaVe08} classification, \object{GRB\,060908} is a clear example of afterglow that is decaying since the first observation. The authors suggest that, in a scenario of a structured outflow observed from different locations, this class of optical light curves could correspond
to an observer location within the aperture of the brighter outflow core, with higher Lorentz factor and, therefore, a shorter deceleration time-scale. We can check whether \object{GRB\,060908} is consistent with the peak flux - peak time correlation found for initially rising afterglows. Our first observation is at $t \sim 61$\,s. The corresponding flux emission at 2\,eV predicted by Eq.\,2 of \citet[][scaled to the redshift of \object{GRB\,060908}]{PaVe08}, assuming we detected the peak optical flux, is  $f_{\rm p} \sim 560$\,mJy ($R \sim 9.4$). The dereddened 2\,eV observed flux is $f \sim 8.2$\,mJy, i.e. substantially lower. The light-curve peak might have occurred earlier than our first observation but things do not improve since the peak flux/peak time correlation is steeper than the observed initial power-law index $\alpha \sim 1.3-1.5$ (Table\,\ref{tab:optfit}). Assuming, for example, that the peak time for the optical light-curve is coincident with the last peak of the prompt emission, $t_{\rm p} \sim 23$\,s, the predicted peak flux would be about only 10 times fainter than that of the extreme \object{GRB\,080319B} \citep{Rac08}. On the other hand, one could attribute the initial steep decay to the reverse shock emission, so that the peak time of the forward shock could be as late as $\sim 365$\,s and be hidden below the reverse shock. In this case, \object{GRB\,060908} is marginally consistent with the relation, although this interpretation requires fine tuning in that the afterglow peak should coincide with the time when the reverse shock emission is no longer dominant. This results therefore suggests that the relation proposed by  \citet{PaVe08} has more scatter than claimed when introducing afterglows which have too early peak to be caught (or, at least, \object{GRB\,060908} is an outlier). The afterglow of \object{GRB\,060908} is also fainter by an order of magnitude than indicated by Eq.\,3 in \citet{PaVe08} at $t \sim 1$\,ks, although this comparison relies also on the amount of the adopted intrinsic extinction (see also Sect.\,\ref{sec:optnir}). There is therefore a clear interest in performing the same check on more GRB afterglows which are already decaying at the time of their first early detection \citep[see e.g. ][]{Kann09}.

The rather long ($>> T_{90}$) temporal interval between the main prompt emission phases and the first afterglow observations makes it unlikely that the initial steeper decay ($\alpha \sim 1.2-1.7$, Table\,\ref{tab:optfit}) is related to the prompt emission. It could consist of the final stages of reverse-shock emission if we assume that we could not detect the predicted faster decay or spectral variation \citep{SaPi99,KoZh07} due to the late observation. If this were the case, following the discussion in \citet{Gomb08}, the tail of the reverse-shock decay would follow a power-law slope of $\alpha_{\rm rs} = (3p+1)/4$, where the electron distribution is assumed to follow a power-law with index $p$ ($dn/d\gamma_e \propto \gamma_e^{-p}$, where $\gamma_e$ is the electron Lorentz factor). With data in Table\,\ref{tab:optfit} this corresponds to $p_{\rm optNIR, 2PL} = 1.64^{+0.32}_{-0.33}$ or $p_{\rm optNIR, 3PL} = 1.43^{+0.25}_{-0.18}$. A $p$ value below $2$ would require a break in the distribution at high energies in order to keep the total energy of the distribution finite. The case for an afterglow characterised by a hard electron distribution index was extensively studied by several authors \citep{DaCh01,Pan05,ReBa08} although numerical and analytical simulations appear to prefer a ``universal" value $p \simeq 2.2$ for particle shock acceleration \citep{Acht01,Vie03}. 

At variance with the expectations from the pre-\textit{Swift} era, most \textit{Swift} GRB afterglows indeed do not show reverse shock emission \citep{Rom06}. Based on the already decaying phase of the afterglow at about 1\,min after the GRB, we can derive a rough estimate of the initial Lorentz factor as $\Gamma_0 \sim 500$ under the same conditions discussed earlier. This estimate 
is in agreement with that based on the \citet{Lia09} $E_{\rm iso}-\Gamma_0$ relation (Sect.\,\ref{sec:pe}) if we assume that the early afterglow is a superposition of reverse-shock decay and forward-shock afterglow onset occurring around the flattening time of the ligth-curve or slightly before.

The steep to shallow transition in the optical resembles the behaviour seen, among others, for \object{GRB\,021211, GRB\,061126} and \object{GRB\,090102} \citep{Li03,Fox03,Kann06,Gomb08,Perl08,Gend09} but it could be due also to a change in the surrounding medium density profile such as that at the termination shock \citep{RamRui01,Chev04,Jin09}.

\subsection{The late afterglow}

The shallow decay which began after 100-400\,s from the burst could be the regular afterglow phase. In this phase, for both a constant density circumburst medium and wind-shaped medium, the difference between the optical/NIR (Table\,\ref{tab:optfit}) and X-ray (Table\,\ref{tab:XRTspec}) spectral slopes suggests that a break frequency is located in between the two bands. If the cooling frequency is located between the two bands \citep[][and references therein]{ZhMe04}, then the spectral slopes should differ exactly by $0.5$. This is consistent in the most favourable case with the observed data only at 2$\sigma$ level since, assuming a power-law electron distribution, they would require for the X-rays $p_{X} = 2\beta_{\rm X} = 2.34^{+0.50}_{-0.44}$ and in the optical a much harder electron spectrum with $p_{\rm optNIR} = 2\beta_{\rm optNIR}+1$ i.e. $p_{\rm optNIR, 2PL} = 1.34^{+0.68}_{-0.80}$ or $p_{\rm optNIR, 3PL} = 1.66^{+0.50}_{-0.58}$. The values of $p$ for the late afterglow are consistent with those derived for the early afterglow within the hypothesis that the early steeper decay is just the tail of the reverse-shock emission.

In the ``slow cooling" phase, afterglows described by a flat electron distribution index are characterised by shallower temporal decays than for softer electron distribution indices, in qualitative agreement with what is observed for \object{GRB\,060908}. The expected decays below and above the cooling frequency, optical/NIR and X-rays bands, respectively, differ by 0.25. In the case of a constant density circumburst medium the higher frequency decays faster than the lower frequencies. The opposite happens for a medium shaped by the wind of a massive progenitor \citep{ZhMe04}.  

It was not possible to strongly constrain the amount of rest-frame dust extinction although in the case of chromatic absorption, correcting for a higher value would generally make the optical spectrum bluer. The SMC extinction curve gave consistently better fits than other curves we tried (Sect.\,\ref{sec:optnir}). Moreover, at the redshift of \object{GRB\,060908}, $z \sim 1.88$, the prominent bump at $2175$\,\AA, which is typical of the Milky Way extinction curve \citep{Pei92}, falls in the $V$ band and therefore its presence could well be probed by our data. In the X-rays, the observed absorption requires additional contribution from the medium surrounding the GRB site in addition to the Galactic one. This contribution is $N_H(z)=8.3^{+5.7}_{-3.7}\times 10^{21}$ cm$^{-2}$. Assuming absorption characteristics similar to those of our Galaxy this would imply an optical absorption of $A_V\sim 5$ mag, however GRB sites are often characterized by much lower optical absorption than that inferred from the X--rays \citep{Str04,Wat07,Camp10}. \citet{Oat09} also reported a low rest-frame extinction for this event from analysis of \textit{Swift} data, although with a redshift implying $(1+z)$ $\sim 20$\% higher than the revised value reported in \citet{Fyn09}, which of course affected their analysis.

Considering the X-ray and optical bands independently of each other, the predicted decay rate in the X-rays would be $\alpha_{\rm X} = (3\beta_{\rm X} -1)/2 = 1.26^{+0.37}_{-0.33}$, i.e. consistent with the observed value. In the optical, the observed decay tends to be too steep unless for instance we assume a wind shaped medium where $\alpha_{\rm optNIR} = (2\beta_{\rm optNIR}+9)/8$, with $\alpha_{\rm optNIR, 2PL} = 1.17^{+0.09}_{-0.10}$ or $\alpha_{\rm optNIR, 3PL} = 1.21^{+0.06}_{-0.07}$, which gives a possible marginal agreement with the observations. The blue, though weakly constrained, optical spectrum would also be consistent with the hypothesis that the optical band is below the injection and cooling frequencies. In this case the spectrum in the optical would be $\beta_{\rm optNIR} = -1/3$ but the decay rate would now be roughly inconsistent with the observations. Also in the ``fast cooling" phase, if the optical band is below the injection frequency but above the cooling frequency, the spectrum is expected to be $\beta_{\rm optNIR} = 0.5$ but again the decay rate would be too shallow. The upper limit at millimetre (Sect.\,\ref{sec:mm}) does not further constrain the afterglow spectrum, since it is roughly compatible with the extrapolation of the optical/NIR spectrum (but for the softest spectra) even without assuming there is a break frequency between the two bands.

The optical/NIR light curve can be modelled with the inclusion of a late steepening which could be either due to the passage of the cooling frequency in the optical/NIR band or perhaps the occurrence of the jet-break. The X-ray light curve, even though statistically does not require such a late-time steepening, can be in agreement with that. In the former case there are two problems. First of all the predicted decay slope ($\alpha \sim 0.9$) is probably too shallow compared to the measured post-transition value (Table\,\ref{tab:optfit}). Moreover, the spectrum after the transition should steepen by $0.5$, as discussed earlier, and although data are not able to strongly constrain the late-time spectral power-law index, this does not seem to be the case. The latter (jet-break) interpretation does not require any spectral evolution and the late-time slope for the $p < 2$ case is predicted to be $\alpha_{\rm jet} = (p+6)/4$, i.e. always steeper than $\alpha \sim 1.5$. This is steeper than the measured value although the late-time slope is based on just a few data points which are likely affected by the contribution of the host galaxy and therefore possibly subject to systematic uncertainties. Following Eq.\,1 in \citet{Ghi06} we can infer a jet opening angle $\theta_{\rm jet} \sim 2^\circ$, a small value but still among those derived for other GRBs \citep{Ghi05}. Knowing the opening angle we can derive the true energy as $E \simeq \theta^2/2~E_{\rm iso} \sim 1.6 \times 10^{49}$\,erg, a value close to the faint end of the observed soft/long GRB energy distribution \citep{Ghi04}. The relatively high brightness of this GRB prompt and afterglow emission \citep{Kann09} would therefore be due to the chance occurrence of observations within the rather narrow aperture cone and with a large bulk Lorentz  factor. However, such a low value for the collimation-corrected energy is essentially inconsistent with the ``Ghirlanda" correlation \citep{Ghi04}. Consistency with the ``Ghirlanda" correlation would require an opening angle larger by about one order of magnitude, corresponding to a jet-break time as late as about 10\,days. The latter would be essentially unobservable in our data set, also owing to the influence of the host galaxy luminosity in the optical/NIR. The disagreement with the ``Ghirlanda" correlation is not by itself a strong argument against the jet-break interpretation of this possible late break. However, it does contribute making this interpretation more contrived \citep[see also][]{McBre10}. 

Finally, we mention that different optical/NIR and X-ray spectral slopes could also result from a more complex electron energy distribution $dn/d\gamma$ than the standard power-law. In particular, the energy distribution of the shock-accelerated electrons may be a broken-power law. For example, $dn/d\gamma_e \propto \gamma_e^{-1.7}$ for $\gamma_m<\gamma_e<\gamma_b$ and $dn/d\gamma_e \propto \gamma_e^{-2.3}$ for $\gamma_e>\gamma_b$, where $\gamma_e$ ($\gamma_m$) is the (minimum) Lorentz factor of electrons accelerated by the shock \citep{PaKu02}. However, whether or not a broken power-law electron energy distribution can account for the current afterglow data depends on the relation between $\gamma_b$ and the dynamics of the forward shock. Unfortunately such a relation is essentially unknown, hampering further investigation of this possibility.

\subsection{The afterglow and the ``cannonball'' scenario}

In the ``cannonball scenario''  the prompt emission is due to the interaction of plasmoids, the cannonballs, ejected by the central engine, with thermal photons upscattered by inverse Compton in a cavity created by the wind blown by the progenitor star or a close companion. The afterglow is instead due to synchrotron radiation from the cannonballs which are sweeping up the ionised circumburst medium \citep[see][and references therein for a comprehensive review]{DaDa09,Dado09}.

Adopting the terminology in \citet{Dado02} and in \citet{Dado09}, the spectral behaviour of an afterglow depends on the location of the so-called bend frequency $\nu_{\rm b}(t)$, i.e. the typical frequency radiated by electrons that enter a cannonball at a given time \citep[see e.g. Eq.\,25 in][]{Dado09}. In the case of an initially wind-shaped medium the bending frequency can be, at early times, well above the optical/NIR bands \citep{Dado07}, the spectrum is expected to be $\beta_{\rm optNIR} \approx 0.5$ and the time decay $\alpha_{\rm optNIR} \approx 1.5$ is in rough agreement with observations (Table\,\ref{tab:optfit}). For the X-ray afterglow the bend frequency is below the X-ray band essentially at all times and the relation $p_{\rm X} = 2\beta_{\rm X} = 2.34^{+0.50}_{-0.44}$ should still hold.  The X-ray temporal decay, however, should be as steep as $\alpha_{\rm X} = \beta_{\rm X} + 1 = 2.17^{+0.25}_{-0.22}$ which appears to be much steeper than the observed value although at early times the data are not able to constrain the X-ray decay index. 

The flattening of the optical light curve could then be interpreted as the transition from a wind-shaped to a constant density environment (much alike within the fireball model) and the X-ray and optical light curves should reach an asymptotic common value of $\alpha = \beta_{X} + 1/2 = 1.67^{+0.25}_{-0.22}$ and $\beta$ in Table\,\ref{tab:XRTspec} for both the X-ray and the optical. As already mentioned for the fireball case, late-time optical data cannot strongly constrain any possible spectral evolution which is however not required by our data. In order to have consistency with the predictions of the cannonball model we should instead assume a late-time evolution of the optical spectrum and a late-time X-ray and optical decay steeper than recorded, possibly hidden because of the inadequacy of the available late-time data and by the contribution of the host galaxy in the optical. This scenario appears somewhat contrived. It is, however, able to coarsely reproduce the overall evolution of the \object{GRB\,060908} afterglow.

\section{Conclusions}
\label{sec:concl}

\object{GRB\,060908} was detected by all \textit{Swift} instruments, securing a large set of observational data for the prompt and the early afterglow phases. Later ground based optical/NIR observations, together with continuous \textit{Swift}-XRT monitoring, allowed us to follow the afterglow evolution for about two weeks and, finally, with observations one year after the GRB, to detect the host galaxy in the $R$ band. The main prompt emission was characterised by two rather broad periods of activity spaced apart by a few seconds of very low emission. A clear correlation between activity and spectral parameters is found as in other cases of GRB prompt emissions. Long lasting high-energy emission for about 1000\,s has also been detected. 

The afterglow light curve in the X-rays is characterised by a continuous decay from the first observation onward. At early times a few relatively small flares are superposed on the decay. The X-ray afterglow is characterised by a synchrotron spectrum generated by an electron population following a rather soft power-law distribution. The optical/NIR light curves show an initial steeper decay, followed by a shallower phase and then by a possible further steepening. The afterglow spectrum is remarkably hard, requiring a flat electron distribution if the emission is modelled by synchrotron emission. Although it is possible to model the optical and X-ray afterglow independently, the multi-wavelength spectral and temporal data challenge available theoretical scenarios.

\object{GRB\,060908} is consistent with the ``Amati" relation while the ``Ghirlanda" relation predicts too late a break to be detected in our data, owing also to the host galaxy contribution flattening the late time light curve decay.

The rich dataset for this event shows that a collaboration among the various teams performing optical/NIR follow-up allowed us to collect data of quality comparable to those provided by \textit{Swift}-XRT, opening the possibility to test GRB afterglow models with much more reliability.

\begin{acknowledgements}
The SMARTS project is supported by NSF-AST 0707627. SC thanks Paolo D'Avanzo, Arnon Dar, Yizhong Fan, Dino Fugazza, Gabriele Ghisellini, Cristiano Guidorzi, Ruben Salvaterra and Zhiping Zin for many useful discussions. AFS acknowledges support from the Spanish MICINN projects AYA2006-14056, Consolider-Ingenio 2007-32022, and from the Generalitat Valenciana project Prometeo 2008/132. We also acknowledge the use of data obtained with the Danish 1.5\,m telescope as part of a program led by Jens Hjorth.
\end{acknowledgements}

\end{document}